\documentclass[a4]{article}
\usepackage{subfigure}
\usepackage{graphicx}

\usepackage{amsmath}
\usepackage{hyperref}

\usepackage{color}
\usepackage[normalem]{ulem}

\newcommand{\Eref}[1]{(\ref{#1})}

\newcommand{\vJ}{{\bf J}}

\newcommand{\vE}{{\bf E}}
\newcommand{\vB}{{\bf B}}

\newcommand{\vH}{{\bf H}}

\begin{document}

\title{{Modeling cross-field demagnetization of superconducting stacks and bulks for up to 100 tapes and 2 million cycles}}

\author{Anang~Dadhich,Enric~Pardo\footnote{Corresponding author: enric.pardo@savba.sk}\\
\normalsize{Institute of Electrical Engineering, Slovak Academy of Sciences,}\\
\normalsize{Bratislava, Slovakia}
}

\maketitle


\begin{abstract}
Superconducting stacks and bulks can act as very strong magnets (more than 17 T), but they lose their magnetization in the presence of alternating (or ripple) transverse magnetic fields, due to the dynamic magneto-resistance. This demagnetization is a major concern for applications requiring high run times, such as motors and generators, where ripple fields are of high amplitude and frequency. We have developed a numerical model based on dynamic magneto-resistance that is much faster than the conventional Power-Law-resistivity model, enabling us to simulate high number of cycles with the same accuracy. We simulate demagnetization behavior of superconducting stacks made of 10-100 tapes for up to 2 million cycles of applied ripple field. We found that for high number of cycles, the trapped field reaches non-zero stationary values for both superconducting bulks and stacks; as long as the ripple field amplitudes are below the parallel penetration field, being determined by the penetration field for a single tape in stacks. Bulks keep substantial stationary values for much higher ripple field amplitudes than the stacks, being relevant for high number of cycles. However, for low number of cycles, stacks lose much less magnetization as compared to bulks. 
\end{abstract}

\flushbottom

\thispagestyle{empty}


\section{Introduction}

{High Temperature Superconductors (HTS) are very promising for next-generation electric devices, such as large-field or large-bore magnets, power-transmission cables, fault-current limiters, transformers, generators, and motors. In rotating machines (motors and generators), HTS enable substantial reductions in weight, size and energy loss compared to their conventional counterparts of the same power and torque. Some examples are wind generators \cite{abrahamsenAB2010SST,lloberasJ2014RER,bergenA2019SST, azo2001}, motors for ship propulsion \cite{snitchlerG2005IEEE, gambleB2011IEEE, zhouD2012SST, haranKS2017SST, namGD2019IES}, and linear motors for high-speed conventional or levitating trains \cite{kusadaS2007IES, unoM2016JRR,  zhengL2011JAP, parkCB2017IEM, gongT2019IES} or Hyperloop \cite{choiSY2019Ene}. } Superconducting motors are also proposed to be used for hybrid distributed electric propulsion for aircraft \cite{asumed2017, haranKS2017SST, bollM2020SST, sugochiR2020IES, grilliF2020JCS}, which can help in reducing CO$_2$ and NO$_x$ emissions by {75~\% percent, fuel burnt by 70~\% and noise by 71 db \cite{haranKS2017SST}}. {REBCO ($RE$Ba$_2$Cu$_3$O$_{7-x}$, where the rare-earth $RE$ is usually Y or Gd)} HTS bulks and stacks {of tapes} can trap high amounts of {magnetic} fields, with current record being 17.6 T for bulks \cite{durrell2014SST} and 17.7 T for stacks \cite{patelA2018SST}, and can be a viable option for building rotors in HTS motors, instead of coils {(using stacks \cite{soteloGG2018IEEE, patelA2018IEEE, smaraA2019SST, grilliF2020JCS} or bulks \cite{zhouD2012SST,haranKS2017SST})}. The main benefits of superconducting stacks or bulks over coils are the compactness of size, sturdy material and geometry, and exclusion of current leads which accounts for substantial reduction of heat load \cite{yanamotoT2018IEEE}. {In addition, stacks of tapes and bulks can also  generate the levitation force in levitating trains and flyweels, see \cite{werfelFN2012SST,soteloGG2015IES,dengZ2016IES} for bulks and \cite{sassF2013IES,patelA2017SST,sassF2018SST,liuK2018SST} for stacks. Finally, stacks of tapes can also enable compact magnets, such as those for Nuclear Magnetic Resonance \cite{hahnS2010IES}. Stacks present several advantages compared to bulks such as higher engineering current densities, longer lengths (up to hundreds of meters), higher engineering current densities, homogeneity, flexibility, and enhanced mechanical and thermal properties thanks to its layered structure \cite{patelA2013SST}.}

{Although stacks and bulks enable to design very powerful superconducting magnets},
they {lose} some magnetization in the presence of alternating {(or AC)} transverse {magnetic} fields, and {hence} the application is left with relatively low power output \cite{campbellAM2014SST,baghdadiM2014APL,baghdadiM2018ScR,kapolkaM2020SST}. The cause for this demagnetization is a well known dissipative effect called dynamic magneto-resistance \cite{brandtEH2002PRL}. 

On application of a transverse AC {magnetic} field above a certain threshold field to a superconducting tape, a DC voltage appears that opposes the present DC transport current. In motors and generators, these transverse AC fields are caused by the background ripple fields, mostly originated by the stator. Then dynamic magneto-resistance or "effective resistance" occurs, which causes the decay of the critical currents and, subsequently, demagnetization of the tape \cite{ogasawaraT1979Cry}. According to Brandt {\emph{et al.}} \cite{brandtEH2002PRL}, this relaxation is due to the "walking-motion" of vortices in the thin tape. However, the dynamic magneto-resistance can also be explained by the Critical State Model \cite{jiangZ2017SST}, and hence the presence of vortices is not strictly required \cite{talantsevEF2018ScR}. Below threshold field, $B_{th}$, a shielded region of "frozen-flux" exists within the center, such that no change in field is experienced by the DC current flowing in this shielded region, and hence the dynamic magneto-resistance is zero because there is no flux crossing the superconductor \cite{jiangZ2017SST}. This resistance is also the cause of an additional power loss that is required to be considered when calculating the total loss in superconductor, which also includes the magnetization loss \cite{ogasawaraT1979Cry}. 

Cross-field demagnetization depends on various parameters, such as the tape geometry and the applied ripple field properties. It increases with the applied field amplitude and frequency, tape thickness, and tape length, and decreases with the critical current density and the width of the tape \cite{mikitikGP2003PRB, campbellAM2017SST, dadhichA2020SST}. The geometry of the HTS stacks also {plays} an important role. Firstly, the cross-field demagnetization rate is roughly inversely proportional to the number of tapes \cite{campbellAM2017SST,baghdadiM2014APL,dadhichA2020SST}. Secondly, ferromagnetic materials can also decrease the demagnetization rate, either by intercalating layers between the tapes \cite{baghdadiM2014APL}, or by placing the stack between ferromagnetic plates \cite{dadhichA2020SST}. More sophisticated structures, such as buried C-shaped stacks in the rotor iron, are also expected to present low cross-field demagnetization \cite{climentealarconV2018JCP}. In the presence of high background ripple field amplitudes and frequencies ranging in a few thousand Hz, as is the case of fully superconducting motors in aviation, the value of dynamic resistance is relatively high, and thus there is a fast decay of magnetization in both superconducting stacks \cite{baskysA2018SST}, and bulks \cite{baghdadiM2014APL, vanderbemdenP2007PRB, kapolkaM2018IES, srpcicJ2019SST, srpcicJ2018IES}. Thus, understanding cross field demagnetization (caused by dynamic magneto-resistance) for superconducting stacks and bulks is an issue of utmost importance. 

Brandt {\emph{et al.}} \cite{brandtEH2002PRL} also {show} that for applied ripple fields below threshold field, the trapped field for a single tape reaches an asymptotic value after high number of cycles that remains indefinitely. It is important to know if the same behavior exists for bulks and stacks, i.e. under what conditions {there will be a non-zero} asymptotic trapped field output, if any{, and its magnitude}. Unfortunately, it has been a very tough and time-consuming task to numerically model high number of cycles with high mesh and field accuracy. Therefore, it is the goal of this paper to model cross field demagnetization for high number of cycles (up to millions) using a unique approach involving dynamic magneto-resistance. Although it has not been currently possible to model stacks involving more than 20 tapes, this paper presents results for up to 100-tape thick stacks, as realistic stacks of tapes often include between 50-100 HTS tapes \cite{climentealarconV2018JCP}. A detailed comparison of superconducting stacks and bulks facing the issue of demagnetization is also required, which this paper achieves.

{The structure of this paper is the following}. Firstly, we present the numerical model developed by our team using dynamic magneto-resistance (DMR model), and give a brief discussion on its benefits. Next, the DMR model is benchmarked with traditional models. Then, the results are presented for a commercial single tape, and stacks involving 10 and 100 tapes, for high number of cycles. We conclude our paper by comparing the potential benefits of stacks and bulks over each other for different cases related to cross field demagnetization.


\section{Modeling Method}

We use our in-house 2D Minimum Electro Magnetic Entropy Production (MEMEP) variational method \cite{pardoE2015SST,pardoE2017JCP} to model the demagnetization behavior of superconducting stacks in the presence of cross-fields. {In 2D,} the ${\bf{J}}$-formulation used in MEMEP is much faster than conventional finite element methods (FEM), since we do not have to consider surrounding air into the mesh, which saves many degrees of freedom \cite{grilliF2010SST}. This method has been extensively benchmarked against conventional $H$-formulation method (COMSOL) and experiments, and have produced excellent results for the given case \cite{pardoE2012SSTa,pardoE2015SST,kapolkaM2018IES,kapolkaM2020SST}. We also have seen that, whereas the 3D method is important for short samples, the 2D cross-sectional simplification can be used when the tape length is 3 times or more than the tape width, producing practically the same results as the full 3D method, but with much less computing time \cite{dadhichA2020SST}.

The MEMEP method can take any $E(J)$ relation into account{, and in particular those described in the following section}. For the primary results of this paper, we use an effective constitutive relation for the dynamic magneto-resistance (DMR model), as described below.

\subsection{Constitutive relations of the superconductor}

{For our 2D cross-sectional approach, where the tapes are infinitely long in the $z$ direction, both electric field, $\vE$, and current density, $\vJ$, follow $\vE=E{\bf e}_z$ and $\vJ=J{\bf e}_z$, being ${\bf e}_z$ the unit vector along the $z$ axis. In this article, we refer to both the magnetic flux density, $\vB$, and magnetic field, $\vH$, as ``magnetic field" because we do not take magnetic materials into account, and hence $\vB=\mu_0\vH$. For ease of terminology, we refer to the magnetic field as simply ``field".}

\subsubsection{Power-law $E(J)$ relation}

A usual choice for modeling superconductors is to use the following power-law E(J) relation, which experimentally agrees with voltage-current measurements for $J$ close to $J_c$. We use this model for benchmarking the proposed DMR model.
\begin{equation}
E(J)= E_c \Bigg( \frac{|{J}|}{J_c} \Bigg)^n \frac{J}{|J|}
\label{EJLaw}
\end{equation}
where, $E$, $J$, $E_c$, $J_c$, and $n$ are the electric field, current density, voltage criterion (10$^{-4}$ V/m in our case), critical current density, and power law exponent, respectively.

\subsubsection{Critical State Model}

The Critical State Model (CSM) postulates that any electromotive force induces current with critical current density, $J_c$, in a superconductor \cite{prigozhinL1997IES}. For infinitely long geometries of finite thickness, $|J|$ only takes the values 0 or $J_c$, but that is not the case for general thin films or 3D shapes, where any $|J|\le J_c$ is allowed \cite{andersonPW1962PRL, srpcicJ2018IES, prigozhinL1996EJAM, bossavitA1994IEEE}. For numerical purposes of MEMEP, we consider the shunted CSM \cite{srpcicJ2018IES, bossavitA1994IEEE}. For benchmarking purposes in $Results$ section, we use the general CSM:
\begin{equation}
  |E|=\begin{cases}
    0, & \text{if $|J| \leq J_c$}.\\
    \rm{any}, & \text{if $|J| > J_c$}.
  \end{cases}
\label{bean}
\end{equation}
{For numerical purposes, we approximate the CSM as a piecewise linear $E(J)$ relation as
\begin{equation}
  E(J)=\begin{cases}
    0, & \text{if $|J| \leq J_c$}.\\
    \rho(|J|-J_c)\frac{J}{|J|}, & \text{if $|J| > J_c$},
  \end{cases}
\label{bean_linear}
\end{equation}
where $\rho$ is a very large resistivity, which could have the physical interpretation of a flux-flow resistivity.
}

\subsection{Effective constitutive relation from dynamic magneto-resistance}

\subsubsection{Dynamic magneto-resistance for a slab}

For cross-field demagnetization in infinitely long problems, it is very convenient to use the following effective $E(J)$ relation, based on the Dynamic Magneto-Resistance (DMR) \cite{brandtEH2002PRL}. For a slab in the CSM in alternating parallel applied field, $B_r$, of amplitude $B_{ra}$, this dynamic magneto-resistance for a given transport current $I$ is \cite{ogasawaraT1976Cry,ogasawaraT1979Cry,brandtEH2002PRL,jiangZ2017SST} $R(I)={2fdl}[B_{ra}-B_{\rm th}(I)]/{I_c}$, with $B_{th}$, $B_{ra}$, $f$, $d$, $l$, and $I_c$ being the threshold field, ripple field amplitude, ripple field frequency, tape thickness, tape length, and critical current, respectively. {Since we assume a slab in the CSM under a parallel AC field,} the threshold field is {\cite{brandtEH2002PRL}} $B_{\rm th}(I) = \mu_0 J_c {d}(1-{{|I|}}/{I_c})/2$, where the critical current, $I_c$, follows $I_c=wdJ_c$, and $w$ being the tape width. Then, for a given current $I$, the DC component of the voltage (time average of the voltage in one applied-field cycle) is due to this resistance $V_R(I)=R(I)I$. Since the DC electric field, $E_R$, is the DC voltage per unit length, $l$, and the average current density in the cross-section, $J_{\rm av}$, follows $J_{\rm av}=I/(wd)$, the effective $E(J)$ relation is
\begin{equation}
E_R(J)=E_0 \left [ \frac{B_{ra}}{B_p}-\frac{B_{\rm th}}{B_p}\left ( \frac{J}{J_c} \right ) \right ]\frac{J}{J_c},
\label{Er1}
\end{equation}
where we omit the sub-index ``$\rm av$'' in the current density, for notation simplicity. We wrote the above equation in terms of normalized dimension-less quantities, with $E_0$ being $E_0\equiv \mu_0fd^2J_c$ and $B_p$ being the penetration field of a slab under no transport current, or ``parallel penetration field'',
\begin{equation}
B_p=\mu_0J_cd/2.
\label{Bp}
\end{equation}
With this normalization, the threshold magnetic field can be written as
\begin{equation}
B_{\rm th}\left ( \frac{J}{J_c} \right )=B_p\left ( 1 - \left | \frac{J}{J_c} \right | \right ).
\label{bth}
\end{equation}

\subsubsection{Effective constitutive DMR model}


\begin{figure}
	\centering
	{\includegraphics[trim= 0 0 0 0, clip, width = 12.5 cm]{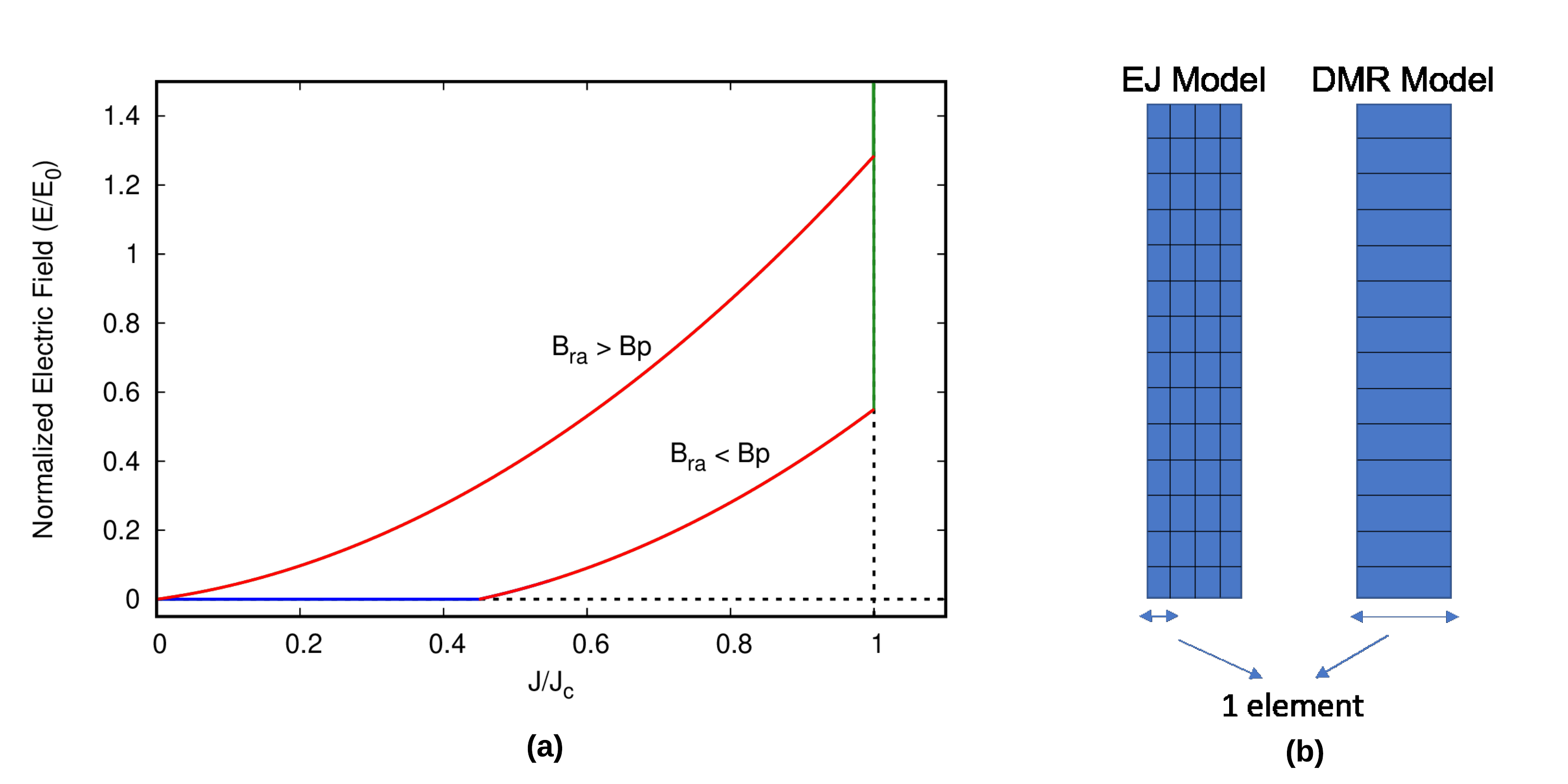}}	
\caption{(a) The E(J) relation developed for the dynamic-magneto resistance (DMR) model. (b) shows that the DMR model only requires 1 element in thickness mesh for the tape which makes the calculations faster. }
\label{EJfig}
\end{figure}

{If we divide the thin film into elements across the width with the same thickness as the tape [figure \ref{EJfig}(b), right], we can approximate each element as an individual slab, thanks to their high aspect ratio and negligible mutual magnetic shielding effects of the parallel component of external fields \cite{brandtEH2002PRL}. Then, we can apply the effective $E(J)$ relation of \Eref{Er1} as local for each element, now being $J$ the average current density a the element, and hence it can vary along the tape width. As already shown for a single tape \cite{brandtEH2002PRL,dadhichA2020SST}, this approach can efficiently model cross-field demagnetization effects in a single tape, as long as it is initially fully saturated due to a perpendicular applied field. Next, we present a more general constitutive relation to also describe demagnetization after partial magnetization or the effect of transverse applied fields simultaneously to the parallel ripple field, as long as their characteristic frequency is much smaller than the ripple field frequency. Although we assume here constant $J_c$, our effective $E(J)$ relation approach can be easily extended for non-uniform $J_c$ across the tape width or a magnetic-field dependent $J_c({\bf B})$.
}

{Combining the CSM $E(J)$ relation of \Eref{bean_linear} and the effective relation of \Eref{Er1} from DMR, we obtain the effective $E(J)$ relation, being}
\begin{equation}
 E(J) = \begin{cases}
	0 & \text{if $B_{ra} \leq B_{\rm th}(J)$ and $|J| \le J_c $}\\
	E_R(J) & \text{if $B_{ra} \geq B_{\rm th}(J)$ and $|J| \leq J_c $}\\
	\rho (|J|-J_c)\frac{J}{|J|} + E_R(J=J_c) & \text{if $|J| \geq J_c$}\\
\end{cases}
\label{Erels}
\end{equation}
{with $E_R(J=J_c)=E_0B_{ra}/B_p=2dfB_{ra}$. As seen in figure \ref{EJfig}(a), for $B_{ra}<B_{\rm th}(J)$ there is no DMR, and hence $E$ vanishes; while for $B_{ra}$ above the parallel penetration field, $B_p$, there is non-zero $E$ for any non-zero $J$.} We hereby use this DMR model in the MEMEP method for finding effects of cross-field demagnetization of superconducting tape and stacks. The benchmarking and discussion for the same can be seen in the $Results$ section. 

\subsubsection{Advantages of DMR model}

There are two major advantages of DMR model over the general $E(J)$ Power Law model. Firstly, the electric field is calculated for the average current density in the tape thickness for the DMR model. Thus, we need only 1 element in the tape thickness, as is shown in figure \ref{EJfig} (b) . In contrast, we need high number of elements ( we use up to 200 elements in mesh thickness) for $E(J)$ Power Law model to accurately see the effects of cross field demagnetization on a HTS tape, which makes the simulations more cumbersome and much slower than the DMR model. Also, the electric field is calculated on an average over one whole cycle of ripple field for DMR model, hence we don't need to consider time steps within each cycle, whereas, for the $E(J)$ Power Law model, high number of time steps are essential for accurate and detailed analysis of change in current density per time step. Thus, due to these reasons, DMR model is much faster when the simulation is to be run for much higher number of cycles (up to millions!), as we present in this paper.

\section{Modeling Configuration}


\begin{figure}
	\centering
	{\includegraphics[trim= 0 0 0 0, clip, width = 12.5 cm]{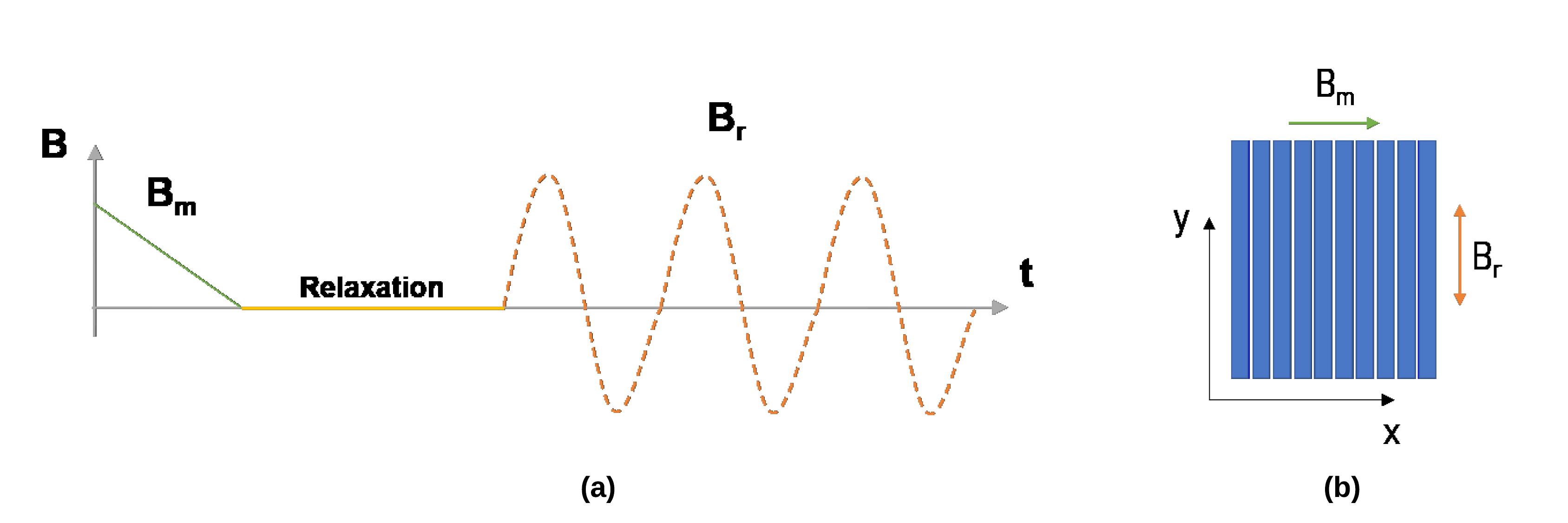}}	
\caption{(a) and (b) show the demagnetization process of a {stack of tapes}. First, the magnetizing field ($B_m$) is applied along the y direction for 100 seconds and the tape is let to relax for 900 seconds. Ripple field ($B_r$) is then applied along the width of the tape for 30 to millions of cycles.}
\label{config}
\end{figure}	

The cross field demagnetization behavior of superconducting stacks {of tapes} and bulks have been analyzed in this paper. {For tapes, we only take the superconducting layer into account, and hence we assume that the conductivity and permeability of the other layers is negligible (stabilization, substrate, and buffer layers).}

The benchmarking of DMR model with $E(J)$ Power Law model has been done on a single tape and a 10 tape stack, with each tape measuring 2 $\mu$m thickness and 12 mm width, with 60$\mu$m gap between each tape in the stack. The current density used for this case is 1.36 x 10$^{10}$ A/m$^2$ at 77 K. The mesh used for $E(J)$ Power Law model is 200 elements in tape thickness and 10 in tape width, with very high accuracy parameters, whereas, for the DMR model, we use only 1 element in thickness and 40 in the tape width. The sample is first magnetized using a field cooling process for 100 seconds with magnetizing field amplitude of 300 mT, followed by a relaxation period of 900 seconds. Later, an alternating sinusoidal cross field of different amplitudes (10 mT- 200 mT) is applied at 500 Hz frequency for 30-100 cycles (Fig. \ref{config} (a) and (b)). 

The stacks used for further analysis are made up of 10 and 100 tapes of 1.5 $\mu$m thickness and 40 mm width, with gap as 100 $\mu$m between tapes. This configuration corresponds to the widest commercially available tape, which is very promising to trap high magnetic fields \cite{climentealarconV2018JCP}. The current density used for this analysis is 5.78 $\times$ 10$^{10}$ A/m$^2$, which roughly corresponds to the values of American Superconductor tape at 30 K. The initial magnetizing field is 5 T for 10 tape stack, and 10 T for 100 tape stack, to ensure the complete penetration and magnetization of the stacks. {We also consider that various amplitudes of cross field (15 mT - 200 mT) are applied at 2400 Hz frequency.} The reason for using such high frequencies in both the cases is to simulate the environment of a superconducting motor, where background ripple fields are at frequencies of this order of magnitude due to the present harmonics. The modeling results in this article used a computer with an Intel i7-7700 processor of 8 logical cores, 16 GB RAM, and Linux operating system.


\section{Results and Discussion}
\label{results}


\subsection{Benchmark of the DMR model}

\begin{figure}[!tbp]
  \centering
	{\includegraphics[trim= 0 0 0 0, clip, width = 12 cm]{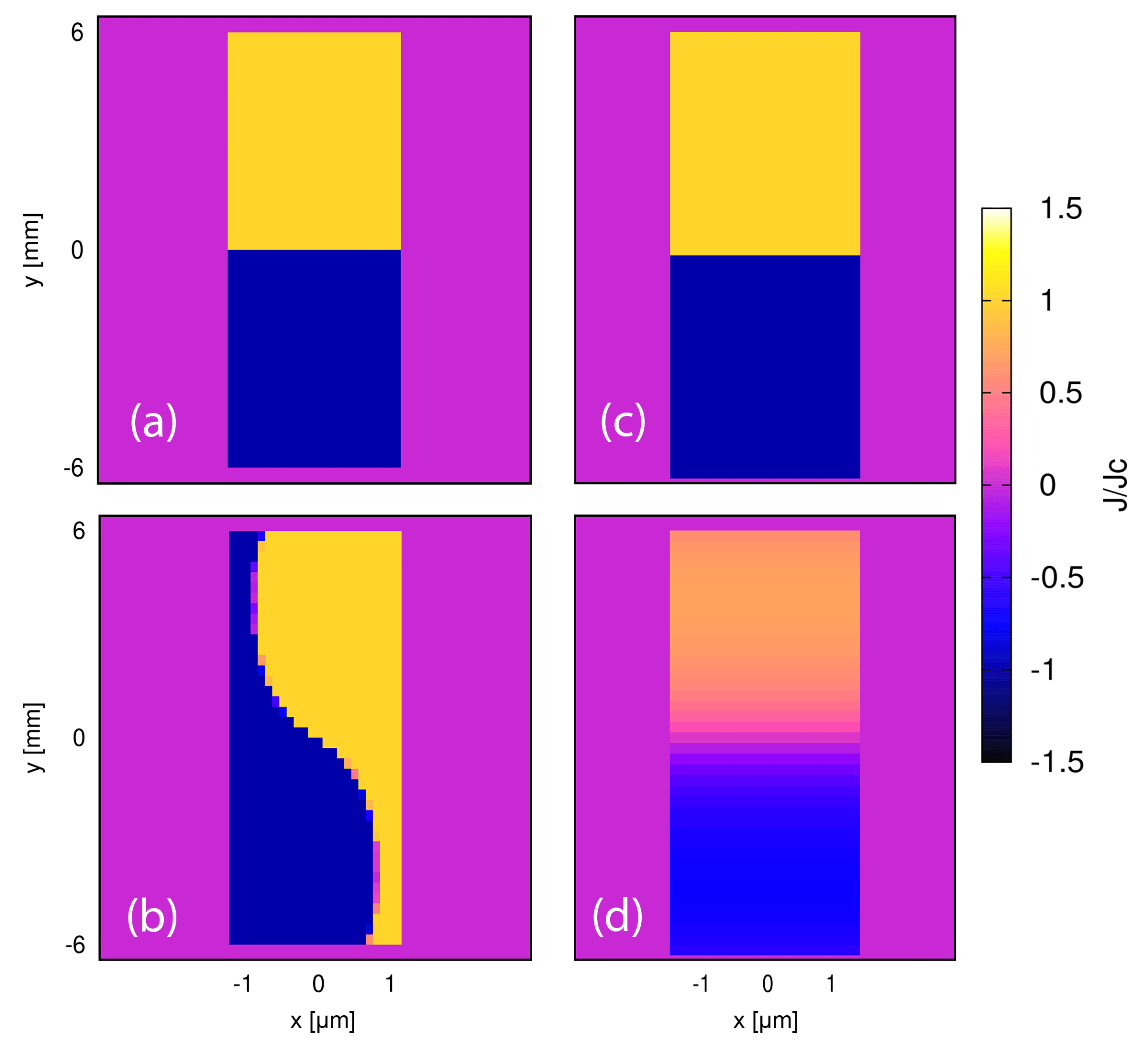}}
\caption{Current density profiles for a single tape under Critical State model(a,b) and DMR model(c,d) after Field Cool Magnetization and Relaxation (a,c), and 30 cycles of demagnetizing ripple field (b,d). Single element is used in the thickness for DMR model. Thickness is also artificially expanded for better analysis.}	
\label{DMR_curprof}
\end{figure}

Firstly, the current density profiles for a single tape using Critical State model and DMR model can be observed in figure \ref{DMR_curprof}. In both cases the tape is fully magnetized at the end of the field cooling magnetization, and we also do not see any relaxation after the magnetization since we have used Critical State Model in the $E(J)$ relations for DMR model. The tape is seen to show considerable demagnetization after application of 30 cycles of ripple field, and we get the typical S shape of the current profile here. For DMR model, the change is only seen in the width of the tape, since we use only single element in the tape thickness (Fig. \ref{DMR_curprof} (d)), in contrast to Critical State Model, where we have up to 24 elements in the tape thickness (Fig. \ref{DMR_curprof} (b)).

\begin{figure}
	\centering	
	{\includegraphics[trim= 0 0 0 0, clip, width = 12.5 cm]{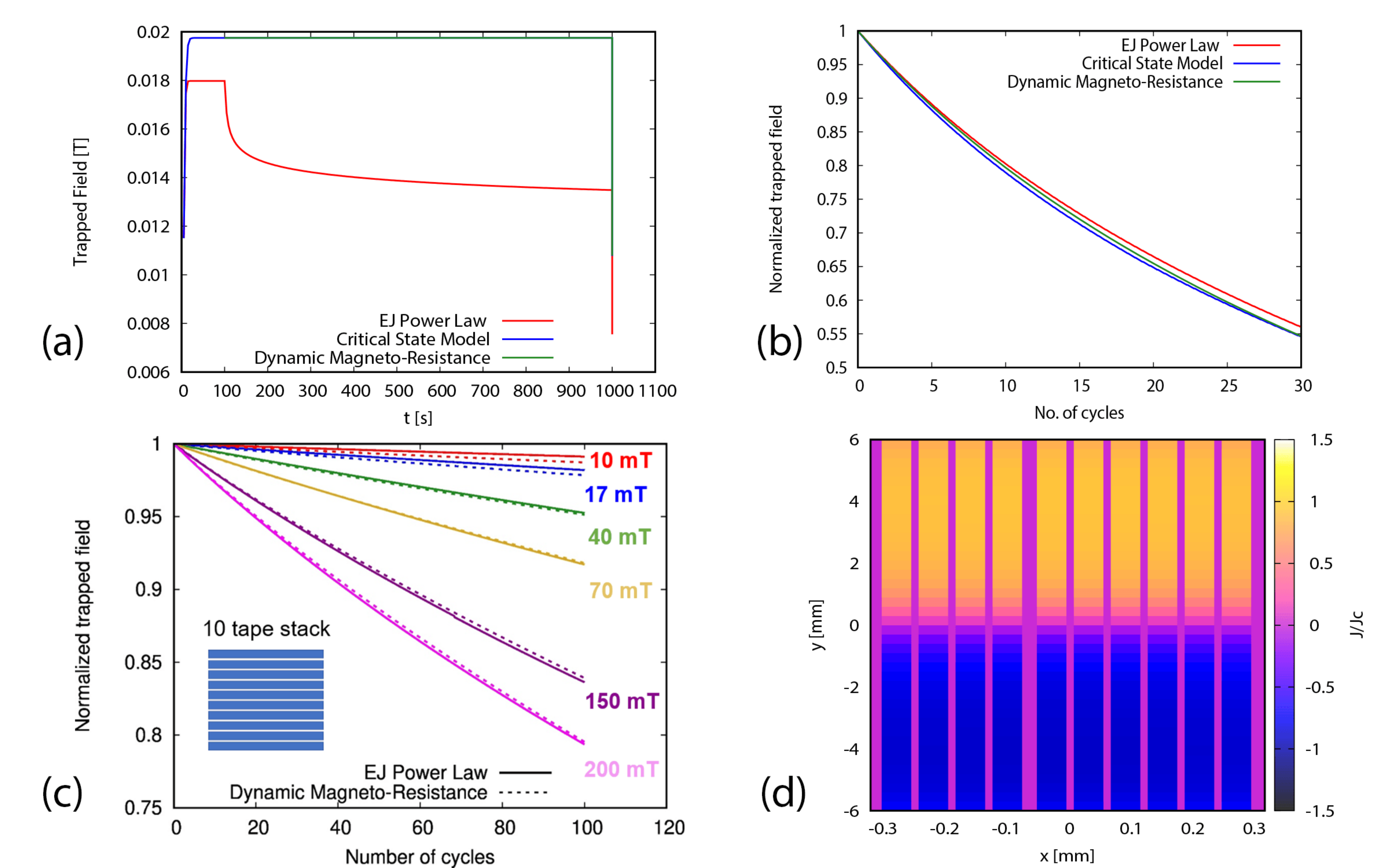}}

\caption{Benchmark of DMR model, and approximated Critical State Model, with $E(J)$ Power Law model for a single tape for (a) the whole trapped field evolution, and (b) the demagnetization process, at 30 cycles of ripple field at 200 mT at high temperature (77 K). A 10 tape stack is compared in (c) for higher number of cycles. DMR model shows good agreement with $E(J)$ power law model for HTS single tape and stack, and is much faster. (d) Current density profile for the stack after 100 cycles of demagnetization at 200 mT using DMR model. Thickness is artificially expanded for better visibility and considerable demagnetization is observed in the stack. }
\label{EJvsDMR}
\end{figure}

Next, we compare DMR model with $E(J)$ Power Law model and approximated Critical State Model (CSM) for a single tape in figure \ref{EJvsDMR} (a) and (b). It can be seen from figure \ref{EJvsDMR} (a) that there is no relaxation for CSM model and DMR model, in continuation with our discussion regarding current density profiles. Figure \ref{EJvsDMR} (b) shows very good agreement between all three cases regarding trapped field behavior during ripple field application. 
In figure \ref{EJvsDMR} (c), a 10 tape stack at 77 K is compared for DMR and $E(J)$ Power Law model for 100 cycles of ripple field. The current density profile for a 10 tape stack after 100 cycles of demagnetization is shown in figure \ref{EJvsDMR} (d). Here again the DMR model agrees very accurately with the $E(J)$ Power Law model. The time taken to simulate this case for $E(J)$ Power Law model was up to 1-2 months, depending on the ripple field amplitude. In contrast to this, DMR model took less than 2 minutes to achieve the same results, which validates the high accuracy, speed, and strength of this model. It is important to state here that for more detailed analyses, such as calculating absolute current densities at each time step and at each section of the tape, $E(J)$ Power Law model should be used. Now, as the DMR model is {successfully} benchmarked, we move forward to simulate demagnetization for thicker stacks and for high number of cycles. 


\subsection{DMR results for high number of cycles}
\label{DMR_high}

%

\begin{figure}
	\centering
	{\includegraphics[trim= 0 0 0 0, clip, width = 12.5 cm]{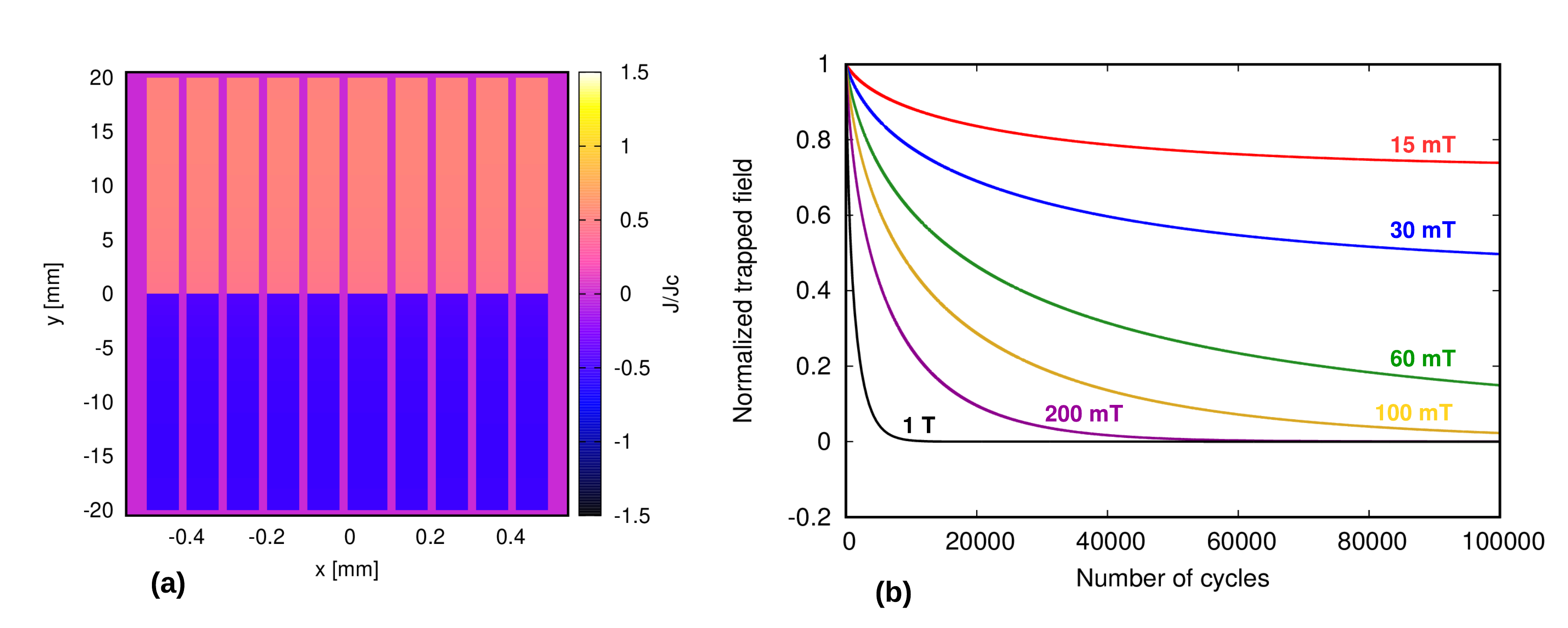}}	
\caption{Demagnetization behavior of a stack of 10 tapes for high number of cycles (100~000) at low temperature (30 K). (a) shows the current profile after 100~000 cycles for 30~mT ripple field. (b) The trapped field reaches an asymptotic value for field amplitudes below the tape's penetration field {(around 55~mT)}. }
\label{100k}
\end{figure}

Firstly, we calculate trapped field behavior for a 10 tape stack at 30 K, with each tape size being 1.5 $\mu$m $\times$ 40 mm, for {100~000} ripple field cycles (see Figure \ref{100k}). It is seen that there is more demagnetization at high ripple field amplitudes, which is due to the dynamic magneto-resistance that exists above the threshold field of the tape \cite{brandtEH2002PRL}. The trapped field goes to zero for ripple field amplitudes above the parallel penetration field of one tape (equation (\ref{Bp})), which is around 55 mT for this case. However, at higher number of cycles, we also see that the trapped field reaches an asymptotic value for ripple field amplitudes below the parallel penetration field. This non-zero asymptotic value appears because for ripple fields below the penetration field, the dynamic magneto-resistance vanishes below a certain current density (Fig. \ref{EJfig}), and hence the current density does not further decay when it reaches this threshold value. With decreasing the ripple field amplitude, the threshold current density increases, also increasing the asymptotic trapped field. This behavior qualitatively {agrees} with Brandt's results for a single tape \cite{brandtEH2002PRL}.

\begin{figure}
	\centering
	{\includegraphics[trim= 0 0 0 0, clip, width = 12.5 cm]{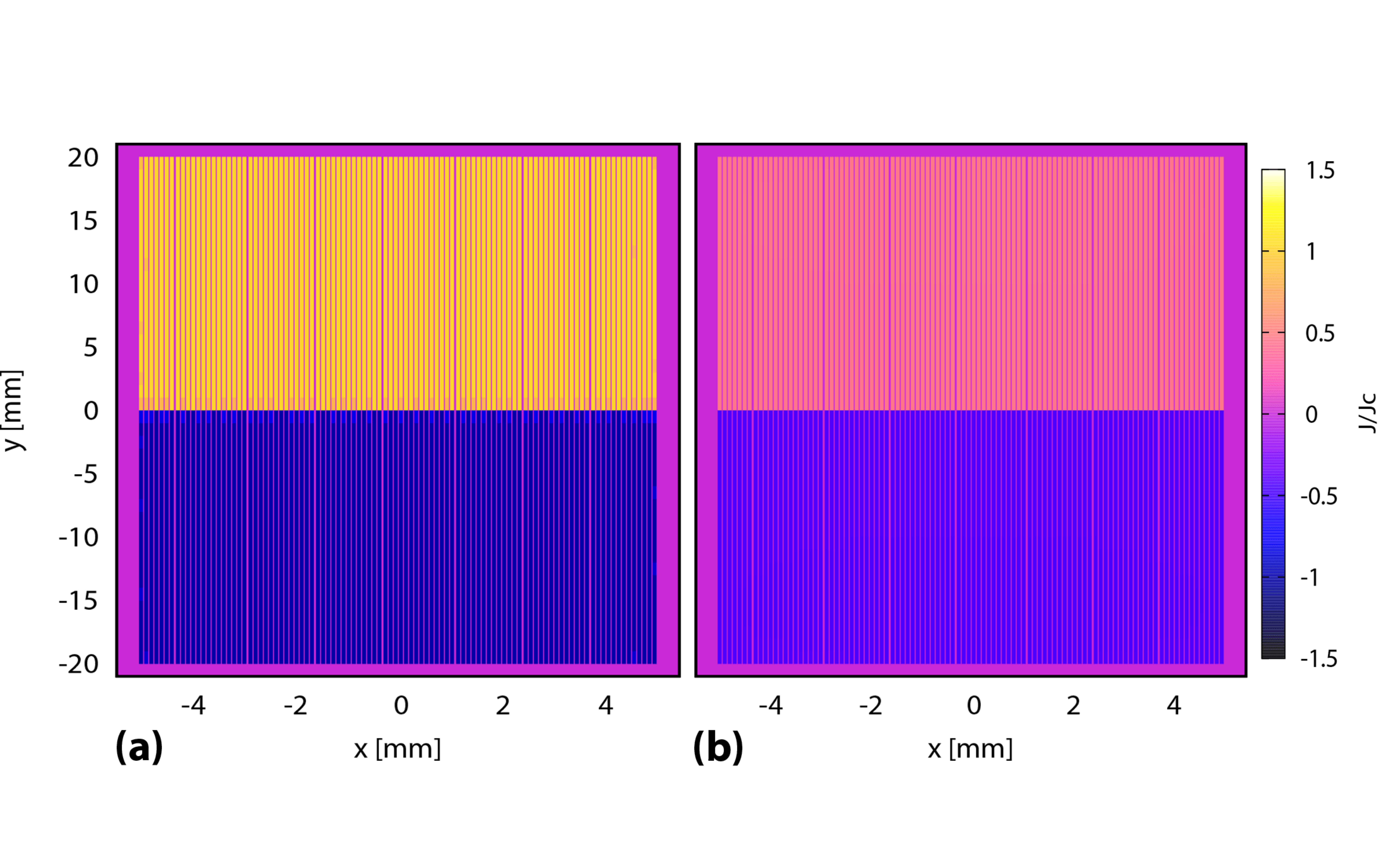}}	
	\caption{Current profiles of a stack of 100 tapes after (a) magnetization and relaxation period, and (b) 2 million cycles of ripple field at 30 mT at low temperature (30 K). Thickness of the tapes is expanded artificially for better visibility of results. At 30 mT, the stack retains almost half of its original magnetization indefinitely.}
	\label{100tapes_curpro}
\end{figure}


\begin{figure}
	\centering
	{\includegraphics[trim= 0 0 0 0, clip, width = 12.5 cm]{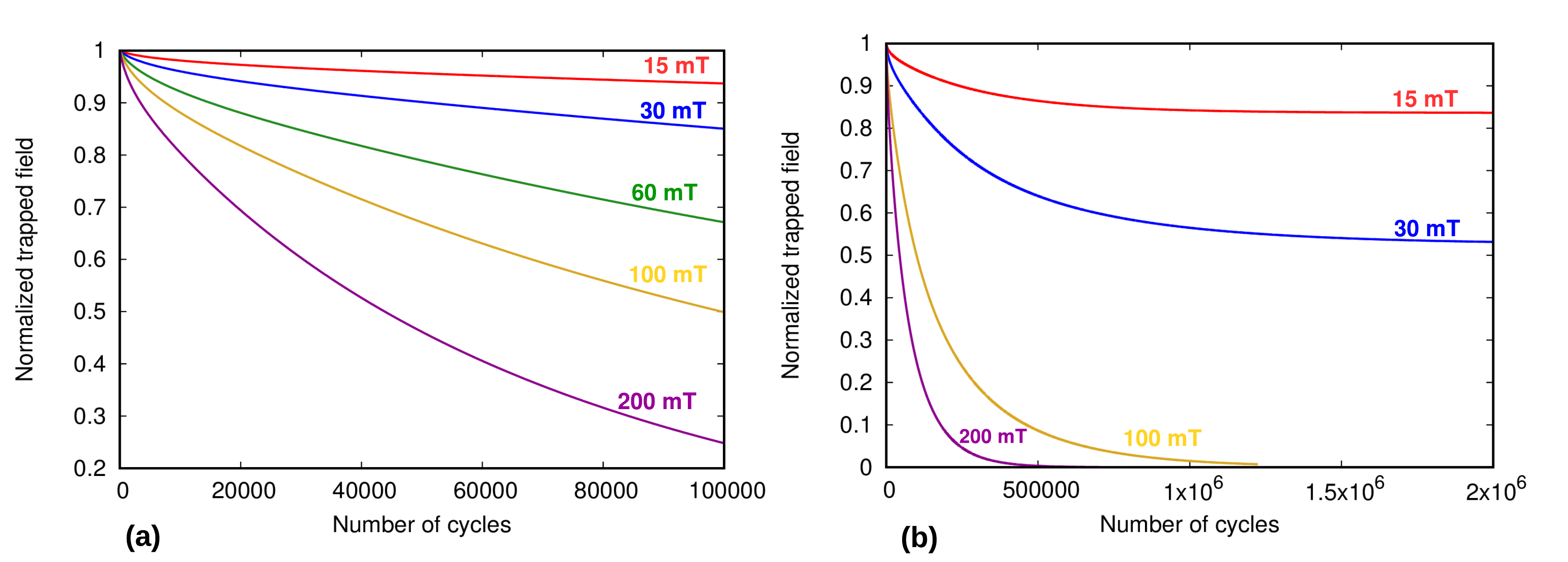}}	
\caption{Demagnetization behavior of a stack of 100 tapes for (a) 100 thousand cycles , and (b) 2 million cycles at low temperature (30 K). The stack reaches an asymptotic value at ripple fields below penetration field of a tape (~ 55 mT). }
\label{100tapes}
\end{figure}

To test the robustness of the DMR model, we also find results for a 100 tape stack for up to 2 million cycles of applied ripple field. The current profiles for this case {are} shown in figure \ref{100tapes_curpro}. It can be seen here that in the case of 30 mT, {the current density in the stack is only reduced by around half of its original value at relaxation, and hence the magnetization is also half} (Fig. \ref{100tapes_curpro}). As for the case of 10 tapes, the reason for this is that at fields below penetration field of the stack (around 55 mT), the trapped field reaches at an asymptotic value, and thus the stack retains considerable amount of magnetization indefinitely. An important point to note here is that the relative retention of the trapped field is independent on the number of tapes in the stack, being the same for 10 and 100 tapes. The reason is that the asymptotic value of the current density only depends on the ripple field amplitude and the parallel penetration field of the tape. Increasing the number of tapes only increases the number of cycles necessary to reach the asymptotic value. 

Figure \ref{100tapes} shows that it is essential to have analysis of very high number of cycles to get a complete understanding of the process. From high number of cycles, we can get answers to questions, such as when exactly the stack will fully demagnetize or when will it arrive at an asymptotic value. For applications such as fully superconducting electric motors for aviation propulsion systems, this is a very important analysis as it can tell us about the eventual power output of the motor after it starts to get demagnetized, and by what time the motor will reach at this range of power. For this case, it is important to note here that two million cycles for such a high frequency (2400 Hz) corresponds to only about 14 minutes. Thus, for example, the magnetization of a superconducting stack with these parameters, in a practical environment, will reduce to approximately 53 percent in just around 7 minutes if the ripple field amplitude is kept to about 30 mT. For lower frequencies, the decay time will be longer. Thus, the primary goal of a motor designer can be to keep the ripple field amplitude as low as possible, and use tapes with the highest available critical current per unit tape width (at the opeartional temperature and background magnetic field) to avoid as much cross field demagnetization as possible. Frequency reduction and increase of number of tapes is secondary, since it only increases the transient decay time.

The time taken for the simulation of 100-tape stack for 2 million cycles is only around 2 days, which affirms to the potential capability of the DMR model. {Moreover, our modeling tool enables to run several calculations in parallel, further reducing the average computing time.}


\subsection{Comparison of bulk vs stack}


\begin{figure}
	\centering
	{\includegraphics[trim= 0 0 0 0, clip, width = 12.5 cm]{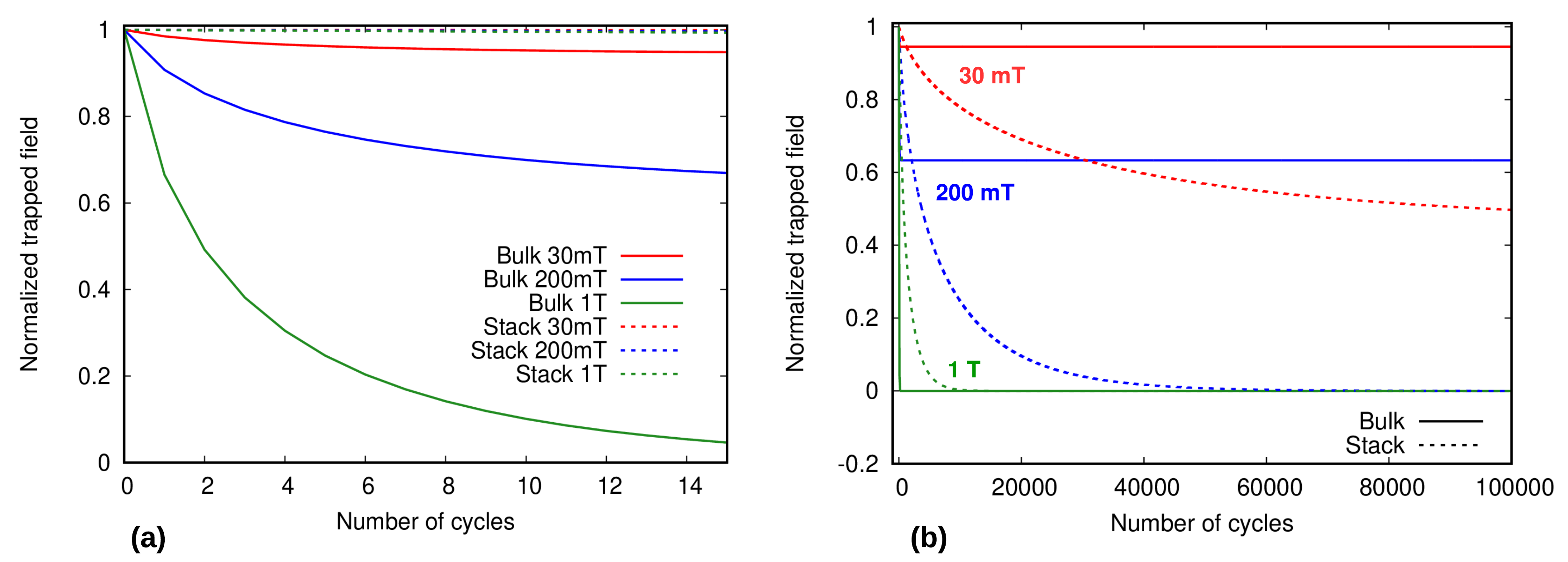}}	
\caption{Comparison of the demagnetization behavior of a bulk and stack for (a) low number of cycles , and (b) high number of cycles. }
\label{bvs}
\end{figure}

\begin{figure}
	\centering
	{\includegraphics[trim= 0 0 0 0, clip, width = 10 cm]{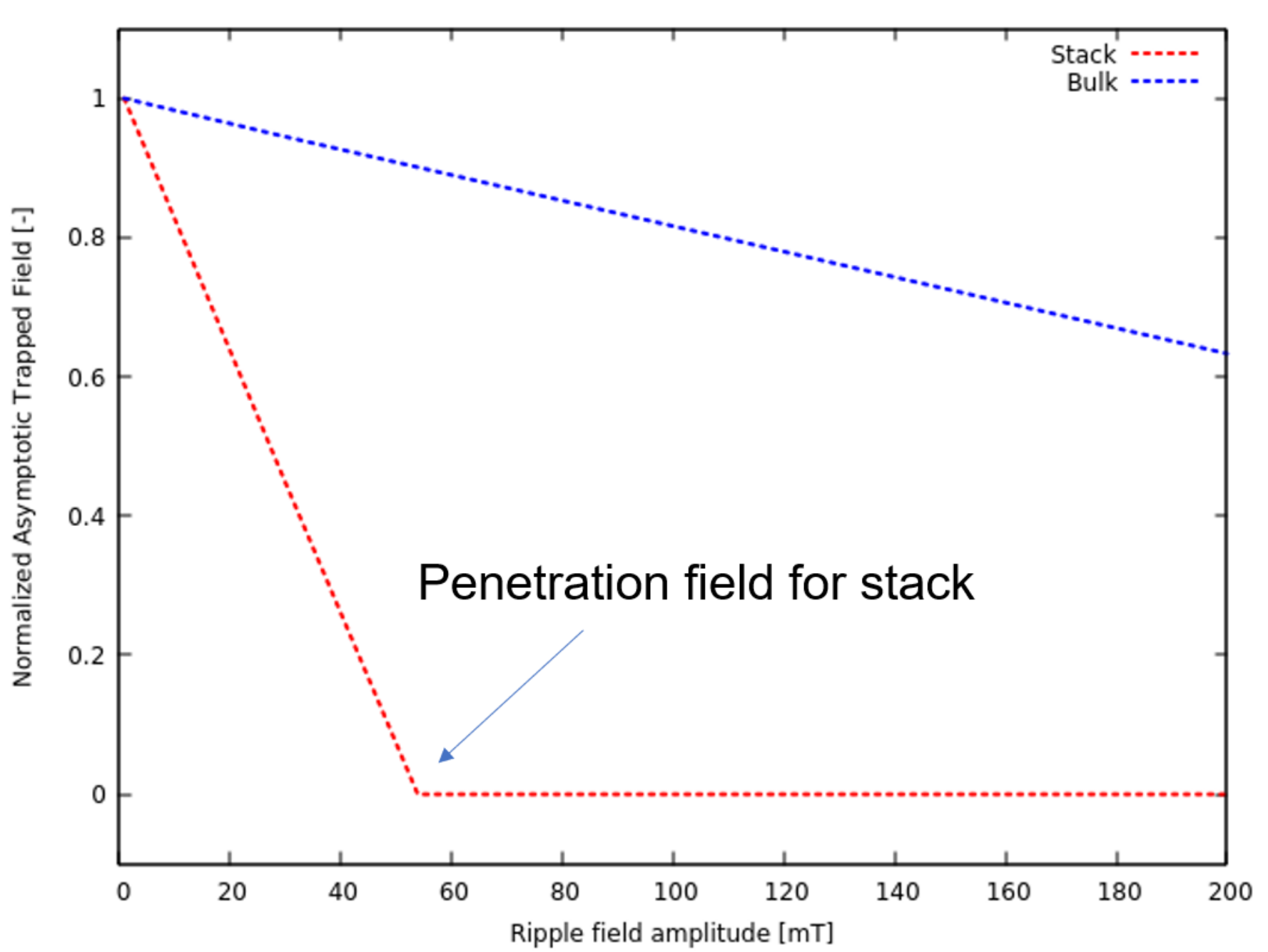}}	
\caption{Asymptotic field values for bulk and stack. Bulk has higher range of asymptotic values as compared to a same size stack. }
\label{bvs_asy}
\end{figure}

Next, we discuss an interesting comparison of a superconducting stack and bulk. For this purpose, we use a 10 tape stack with the same parameters as in previous section (for 30 K). 
{The dimensions of the equivalent bulk used here are the same as the stack; as well as the engineering critical current density (or total current density per unit overall volume), being}
\begin{equation}
{ J_{ce} = J_c n_t d \frac{1}{D} },
\label{Jce}
\end{equation}
where, $n_t$, $d$, and $D$ are number of tapes in the stack, {superconductor layer} thickness of a tape in the stack, and total height of the stack, respectively.

The comparison of the bulk and stack can be seen in the figure \ref{bvs}. The parts (a) and (b) show a striking contrast between both the bulk's and the stack's trapped field behavior during demagnetization. For low number of cycles, we see that the stacks are much better than the bulks at all ripple field amplitudes, with losing almost no magnetization, whereas the bulks show a very fast transient decay for the same cases, and even going to zero at very high ripple field amplitudes. Figure \ref{bvs} (b) shows that although the bulks experience a very fast initial decrease of magnetization, the permanent asymptotic values are much higher than for the stacks. The reason for such a behavior is that the bulks have much higher parallel penetration field than the stacks, which provides them with high range of ripple field amplitudes at which they can attain asymptotic field values (figure \ref{bvs_asy}). Thus, for high number of cycles, the bulks are better, as their stable asymptotic field values are much higher than the stack. For low number of cycles, the stacks are much better than the bulks because stacks require orders of magnitude more cycles to demagnetize. Therefore, for applications with high amplitudes of ripple field and low number of cycles (or very low frequencies), HTS stacks are preferred. In addition, they also have some other benefits over bulks, such as ease and flexibility of construction. However, for applications involving high ripple fields amplitudes and frequencies (or long run times), bulks should be used, because of their higher asymptotic values as compared to {stacks}.


\section{Conclusion}

Cross field demagnetization of stacks is a major issue for a fully superconducting motors. The model based on dynamic magneto-resistance (DMR Model) can very accurately predict this situation for high number of ripple field cycles (up to millions), and very thick stacks, at a very high speed as compared to the conventional numerical models using $E(J)$ Power Law. The results show that, for high number of cycles, superconducting stacks reach a non-zero {static} asymptotic value{,} if the applied ripple field is below the parallel penetration field of a tape. {This asymptotic value depends on the tape properties only, and hence it does not increase with the number of tapes in the stack}. {For bulks, the asymptotic value} is much higher than {for} stacks. We also see that bulks reach asymptotic values {much} faster than the stack. In contrast, stacks do not lose much magnetization if the runtime of the application involve low number of cycles, even if the ripple field amplitudes are high. In motors for aviation, where ripple fields can range from few hundreds of Hertz to kilo-Hertz, the asymptotic value could be reached in times much shorter than the flight-times, being bulks more reliable for these applications{, unless the ripple fields in the stack are strongly reduced \cite{climentealarconV2020JPP}}.



\section*{Acknowledgements (not compulsory)}

The authors acknowledge the financial support by the European Union's Horizon 2020 research innovation program under grant agreement No 7231119 (ASuMED consortium), as well as from the Grant Agency of the Ministry of Education of the Slovak Republic and the Slovak Academy of Sciences (VEGA) under contract No. 2/0097/18.

\section*{Author contributions statement}

Both AD and EP wrote the article; EP conceived the numerical simulations background and physical model; AD further developed the simulation software and performed the numerical simulations; EP concluded the analysis and interpretation; AD prepared all figures.

\section*{Additional information}

\textbf{Competing interests:} The authors declare that they have no competing interests. 


\end{document}